\documentclass[preprint,preview,12pt]{elsarticle}

\usepackage{graphics}
\usepackage{graphicx}

\usepackage{amssymb}

\usepackage{latexsym}
\usepackage{makeidx}
\usepackage{multirow}
\usepackage{multicol}
\usepackage{float}
\usepackage{subfigure}
\usepackage{color, colortbl}
\usepackage{xcolor}

\usepackage{rotating}
\usepackage{caption}

\biboptions{square,numbers, comma, sort&compress}

\begin{document}

\begin{frontmatter}



\title{Understanding the input-output relationship of neural networks in the time series forecasting radon levels at Canfranc Underground Laboratory}


\author[mcm]{I\~naki, Rodr\'iguez-Garc\'ia, Miguel C\'ardenas-Montes}
\address[mcm]{CIEMAT, Department of Fundamental Research. \\Avda. Complutense 40. 28040. Madrid, Spain.}
\ead{inaki.rodriguez@ciemat.es, miguel.cardenas@ciemat.es}

\begin{abstract}
Underground physics experiments such as dark matter direct detection need to keep control of the background contribution. Hosting these experiments in underground facilities helps to minimize certain background sources such as the cosmic rays. One of the largest remaining background sources is the radon emanated from the rocks enclosing the research facility. The radon particles could be deposited inside the detectors when they are opened to perform the maintenance operations. Therefore, forecasting the radon levels is a crucial task in an attempt to schedule the maintenance operations when radon level is minimum. In the past, deep learning models have been implemented to forecast the radon time series at the Canfranc Underground Laboratory (LSC), in Spain, with satisfactory results. When forecasting time series, the past values of the time series are taken as input variables. The present work focuses on understanding the relative contribution of these input variables to the predictions generated by neural networks. The results allow us to understand how the predictions of the time series depend on the input variables. These results may be used to build better predictors in the future.
\end{abstract}


\end{frontmatter}


\section{Introduction}
Nowadays, underground Physics experiments are focused on searching particles with low energy signals such as dark matter direct detection. The searched signals are, usually, very weak and with low expected frequency. For these reasons, it is fundamental to control and understand as much as possible the sources of background. For example, for minimizing the contribution of the cosmic rays, these experiments are carried out in subterranean facilities such as mines and caves under mountains. 

These underground facilities are enclosed by rocks in the mountains. The rocks emit radon, towards the caves, which is a potential source of background. In the underground laboratories, the radon is produced by the decay of radium-226 present in the rocks. With an activity ranging from tens to hundreds of $\frac{Bq}{m^3}$, the radon-222 decays into polonium-218 and emits an $\alpha$ particle. Due to the high rate of emission, the $\alpha$ particles can hide the searched signal.

The Canfranc Underground Laboratory (LSC) is a laboratory located in the Spanish side of the Pyrenees. Over the laboratory there are about 800 meters of rocks, that gives natural protection against cosmic rays. For this reason, the LSC is a good host for low-energy events search experiments like ANAIS \cite{Amare:2012ex}, ArDM \cite{Calvo_2017} and DArT \cite{Garcia_2020}. For the international collaborations whose experiments are located at LSC, it is useful to have high-quality predictions of the radon levels in order to schedule the maintenance operations during the low-level periods. Minimizing the exposure of the materials is therefore an objective.

The present paper continues with the forecasting efforts of the radon levels at LSC. In the past, classical time series algorithms such as Holt-Winters, ARIMAs and STL Decomposition have been compared to deep models in \cite{DBLP:conf/hais/Mendez-JimenezM18}. An ensemble of STL Decomposition joined to the Convolutional Networks, called STL+CNN, to improve the forecasts, was presented in \cite{DBLP:conf/caepia/Mendez-JimenezM18} for various datasets and, in \cite{DBLP:conf/softcomp/MontesM19} for the radon Series. Also in \cite{HAIS2020} the authors proposed a population-based incremental learning (PBIL) algorithm to optimize the hyperparameters of STL+CNN models. Other novel research lines have been tested, in \cite{tomas2020} the authors proposed to use the weather variables from four cities around the LSC, to improve the predictions of radon levels. The analysis of the uncertainty in Deep Learning models and Gaussian Process was done in \cite{IGPL2020}. The main goal of the present paper is to understand how the input influences the predictions of the neural networks which forecast the radon-222 at levels. For this task, the algorithms of Garson \cite{garson1991} and Olden \cite{olden2002} are applied. 

The rest of the paper is organized as follows: in Section \ref{section:MM}, the radon time series, the neural networks and the importance algorithms are described. The results and their posterior analysis is shown in Section \ref{section:results}. Finally, Section \ref{section:conclusions} presents the conclusions of this work.

\section{Methods and Materials\label{section:MM}}
\subsection{Canfranc Underground Laboratory}
The LSC is located between a road tunnel and old train tunnel that unites Spain and France across the Pyrenees. The laboratory is composed of three sites (LAB780, LAB2400 and LAB2500) under 850 meters of rocks which give a cosmic rays protection of 2450 m.w.e. The main site is also divided in three halls (A, B and C) for hosting the diverse experiments.

\begin{figure}[!h]
    \centering
    \includegraphics[width=0.9\textwidth]{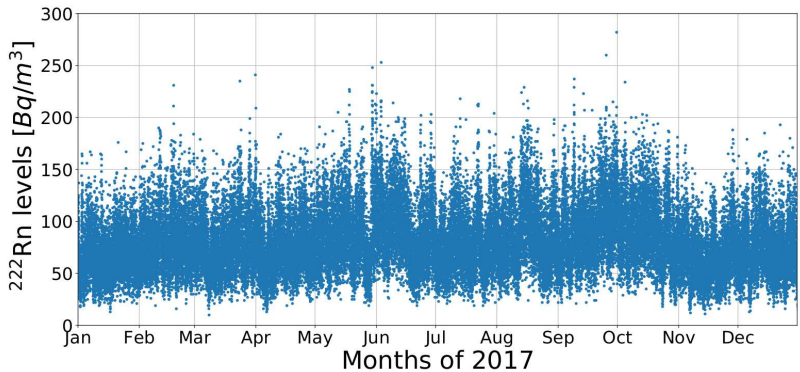}
    \caption{Example of raw measurements of the radon-222 concentration in the Hall A of the LSC collected in 2017.}
    \label{fig:radon_year}
\end{figure}

 In the Hall A, the radon level has been recorded by an Alphaguard P30 every ten minutes since July 2013. The raw measurements have a lot of noise as it can be seen in Figure \ref{fig:radon_year}. Few missing values are in the data set. The large gaps appear in July 2014 with 913 missing values -about a week-, in June 2015 with 1053, and in January 2016 with 585 -about four days.
 
\begin{figure}[!h]
    \centering
    \subfigure[]{\includegraphics[width=.88\textwidth]{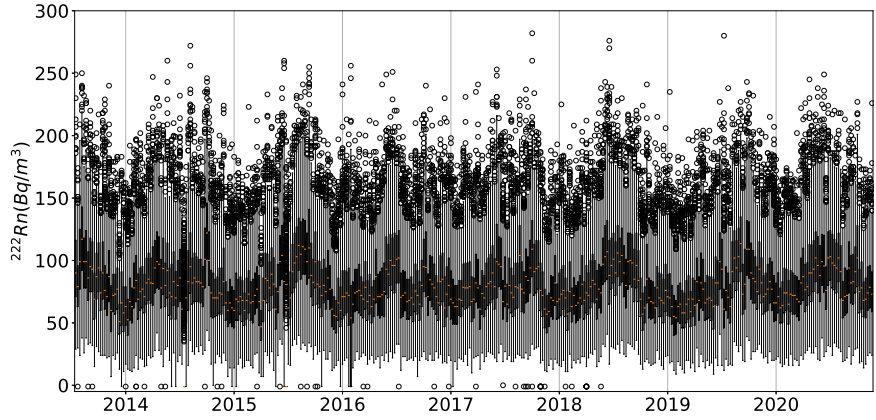}
        \label{fig:rn_box:a}}
    \subfigure[]{\includegraphics[width=1.\textwidth]{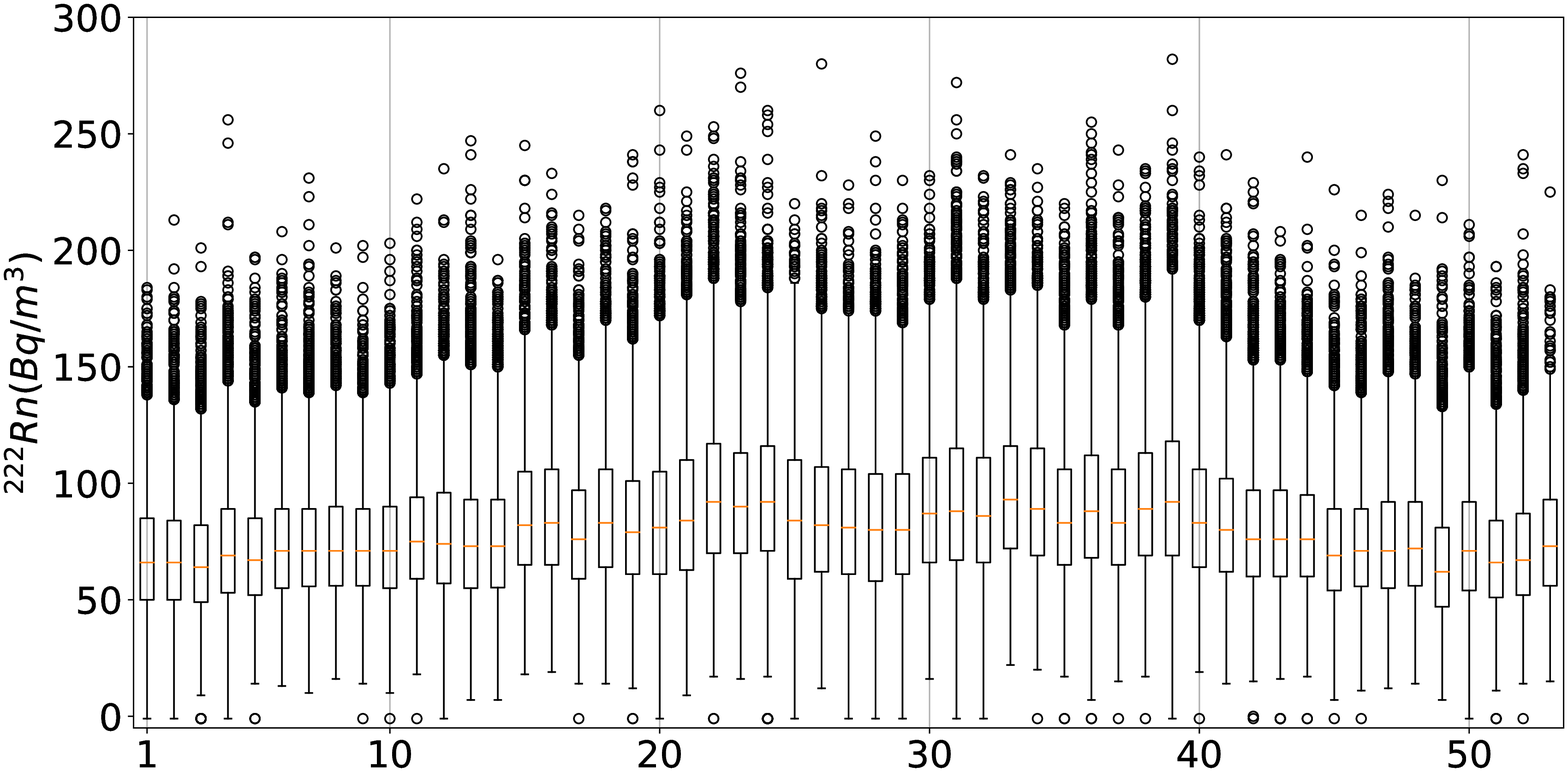}
        \label{fig:rn_box:b}}
    \caption{Weekly box-plots of radon-222 level at Hall A of the LSC, by weekly over range of the years (a) and by week independently the years (b).}
    \label{fig:rn_box}
\end{figure}

 Following previous work \cite{DBLP:conf/hais/Mendez-JimenezM18}, the levels of radon exhibit a certain seasonality in the medians, as Figure \ref{fig:rn_box} shows, where the weekly medians are lower in winter than in the summer. For this reason, the work-data are the weekly medians of the raw values in order to reduce the noisy contributions. After sampling data into weekly values, there are only two missing values. These are filled after a Gaussian distribution over the time series.

\begin{figure}[!h]
    \centering
    \includegraphics[width=.95\textwidth]{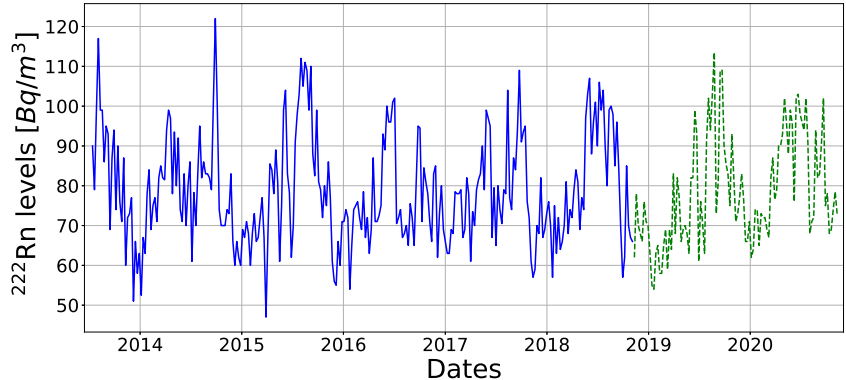}
    \caption{Final dataset composed of the weekly medians of the radon-222 measured in the LSC from July of 2013 to November of 2020. The blue line corresponds to the values of the train set, and the dashed green line to the test set.}
    \label{fig:rn_series}
\end{figure}

The final time series is formed by the 383 median weekly values (orange points in Figure \ref{fig:rn_box:a} from July 21th, 2013 to November 8th, 2020. The last two years (about the 30\%), are reserved to the test set. This implies that the last date of the train set is November 4th, 2018; and the test set begins the next week, on November 11th, 2018. Both data sets are plotted in Figure \ref{fig:rn_series}.

\subsection{Encoding the input}
In machine learning, the dimensions of the input data depend on the nature of the problem being analyzed. When the input data is an univariate time series, the input is a 1D vector, but, for a multivariate time series the input is a 2D vector. When the problem deals with images, the input is 2D matrix or 3D tensors, depending if  the images have one or three channels. Also the input can have more dimensions for examples where several images are concatenated in a sequence of images.

Time series seem as an unique vector or example with size as large as the number of records. The most frequent way of creating the independent and dependent variables in machine learning is to divide the series in several frames or sub-series of a fixed size $L$. The look-back $L$ indicates how much past values are incorporated in the input frames. 

As an example, in the weekly data, if $L=4$, it indicates that the frames are formed by the information of the previous month, but if $L=12$ the frames are formed by the previous quarter.

To show how the time series is divided in frames, the following time series X is used. It is composed of ten values and it will be divided into sub-series with size of seven values.

\begin{equation}
    X = (x_1 , \ x_2 , \ x_3 , \ x_4 , \ x_5 , \ x_6 , \ x_7 , \ x_8 , \ x_9 , \ x_{10}, \ x_{11}) 
\end{equation}
The sub-series are:
\begin{equation}
    \begin{array}{ccc}
      X_1 &= & (x_1 , \ x_2 , \ x_3 , \ x_4 , \ x_5 , \ x_6 , \ x_7)     \\
      X_2 &=  &(  x_2 , \ x_3 , \ x_4 , \ x_5 , \ x_6 , \ x_7,\ x_8)    \\
      X_3 &= & ( x_3 , \ x_4 , \ x_5 , \ x_6 , \ x_7, 
      \ x_8 , \ x_9 )     \\
      X_4 &=& ( x_4 , \ x_5 , \ x_6 , \ x_7, \ x_8 , 
      \ x_9 , \ x_{10})    \\
    \end{array}
\end{equation}
And now, the sub-series $X_1, X_2,X_3$ and $X_4$ can be used to predict the values $ x_8, x_9,x_{10}$ and $x_{11}$ respectively, forming a dataset with four events. 
The number of events in the dataset are the length of the original time series and minus the input window length.
Each variable of the sub-serie is named as $lag_i$ where $i$ is the position in the sub-serie, and the following value is called $actual$. In the example below, $ x_8, x_9,x_{10}$ and $x_{11}$ are values of the variable $actual$ for each event. The $lag_1$ refers to the immediate previous value of $actual$ value, if $actual=x_8$, then $lag_1=x_7$, and $lag_L$ is the last value of the sub-serie; in the same example, $lag_7=x_1$.

\subsection{Neural Networks}
Artificial Neural Networks (ANN) are biological-inspired algorithms based in the brain networks relations. The fundamental unit of ANN, called neuron, is a nonlinear combination of input values and weights as it is shown below (Eq. \ref{hidden}):
 \begin{equation}
a_n =  f_h(b_n+w^{1}_{n}x_1+w^{2}_{n}x_2+w^{3}_n x_3+...+ w_{n}^i x_i+...+ w_{n}^I x_I) \label{hidden}
\end{equation}
where $a_n$ is the value of the neuron $n$, $x^i$ is the i-input, $w_n^i$ is the weight of the neuron $n$ respect the input $i$, $b_n$ is the bias unit of the neuron $n$, and $f_b(\cdot)$ is the activation function of the hidden layer, and it has to be a nonlinear function. The neurons are grouped in layers; where the relationships between layers are: the outputs of one layer are the inputs for the next layer. The layers are divided in three classes; the \textit{input layer} is the first layer of the neuron, the final layer is called \textit{output layer} whose values are the network results (or predictions); the rest of layers are \textit{hidden layers}. In the following only ANN with a single hidden layer is considered.

The equation for the output neuron is similar to Eq. \ref{hidden}, where the input x has to be changed by the hidden neurons (Eq. \ref{output}):
    \begin{equation}
    \hat{y} = f_o(v_0+v^{1}a_1+v^{2}a_2+v^{3} a_3+...+v^{N} a_N+...) \label{output}
\end{equation}
where $\hat{y}$ is the prediction of the neural network, $v^n$ the weight of the output neuron respect the hidden neuron $n$ and $f_o(\cdot)$ is the activation function of the output layer.

\subsection{Evaluating the Input Importance}
This work follows the algorithms developed by Garson \cite{garson1991} and Olden \cite{olden2004} in order to know how the input influences the predictions of the models. 

The Garson’s algorithm was presented in 1991\footnote{The original paper where the Garson's algorithm was presented, is not available. The details can be found in \cite{goh1995, olden2004}.}, it calculates the importance of the input using the absolute value of the products between the weights of the neurons of the hidden layer $w_n^i$ and the weights of neurons of the output layer $v^n$ (Eq. \ref{eq:garson}). 

\begin{equation}
c_i = \sum_{j=1}^m \frac{|w_j^i \cdot v_j|}{\sum_{k=1}^I |w_j^k \cdot v_j|} 
\label{eq:garson}
\end{equation}
where $w$ are the weights of the neurons in the hidden layer, $v$ are the weights of the neurons in the output layer, $m$ are the number of the neurons in the hidden layers, and $I$ are the number of the dimensions of the input variables. Variables with low importance in the prediction obtain coefficients $c_i$ close to zero. 

The Olden's algorithm has a similar structure to Garson's one but skips the normalization. This allows coefficients positive and negative. Variables with low relevance produce coefficients $c_i$ close to zero.  
In the comparison \cite{olden2004} authors claim a higher performance of the former in comparison with the latter one. 

\begin{equation}
c_i = \sum_{j=1}^m w_j^i \cdot v_j
\label{eq:olden}
\end{equation}
where $w$ are the weights of the neurons in the hidden layer, $v$ are the weights of the neurons in the output layer, and $m$ are the number of the neurons in the hidden layers.

\subsection{Models and Training}
Before applying the algorithms previously described, a grid search is implemented to know which hyperparameters configuration produces the best prediction. The architecture of the neural networks
is: one input layer with the same neurons as the window input $L$, followed by one hidden layer which has $N$ neurons with \texttt{relu} as activation function. Then, an output layer with one neuron is applied whose function is \texttt{linear}. The size of the input
and the number of hidden neurons are the hyperparameters being evaluated in the grid search. The size of input ranges between 10 and 60 lags values with step of one, and for the number of neurons
from 10 to 100 with step of 5.

During the train sessions the callback \texttt{EarlyStooping} is used. It permits to stop training after a number of epochs (chosen in 20) when the mean square error does not improve. It is also able to restore the weights of the best epoch. The initial number of epochs was chosen in 2000. The rest of the hyperparameters are:

\begin{itemize}
    \item A batch with a size of 32 events.
    \item The optimizer is Adam with the defaults parameters.
    \item The loss function is the mean squared error (MSE), and also the root mean squared error (RMSE) and the mean absolute error (MAE) are recorded as metrics.
\end{itemize}

During the grid search, ten independent executions of each network are performed, and to discriminate the best model, the RMSE is the figure of merit.

The neural networks were implemented on a NVIDIA Pascal GPU with the use of the library \texttt{Tensorflow} \cite{tensorflow2015-whitepaper}. The libraries  \texttt{scikit-learn} \cite{scikit-learn}, \texttt{pandas}, \texttt{numpy}, \texttt{matplotlib} and \texttt{seaborn} are also used.

\section{Results and Discussion\label{section:results}}
\subsection{Grid Search and Best Performance}
\begin{figure}[!h]
    \centering
    \includegraphics[width=1.\textwidth]{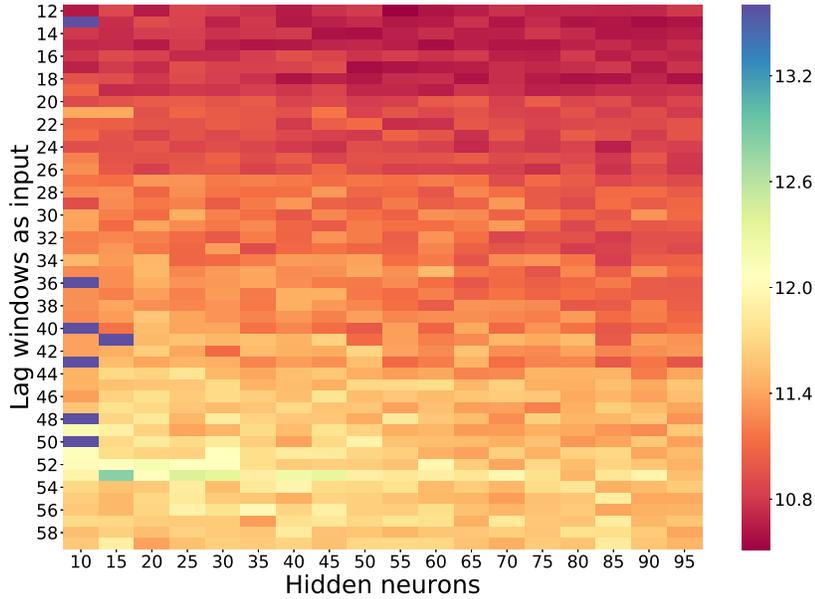}
    \caption{Results from the grid search process after 10 independent runs.}\label{fig:search}
\end{figure}

The best one-layer architecture to forecast the radon-222 level in the LSC is chosen after the grid search process. It searches between two hyperparameters, the input window size ---the number of lags used for forming the input--- and the number of neurons in the single hidden layer. Since the weights of neurons are used for evaluating the relative importance of the variables, they must be as close as possible to the optimal set of weights. The results of the grid search are summarized in Figure \ref{fig:search}.

The metrics for the ten best architectures from the grid search are detailed in Table \ref{tab:results}. The lowest error corresponds to a model with 12 past values as input and 55 hidden neurons. It is also notable that the rest of high-quality configurations use low inputs, in the order of a dozen of lagged values.

\begin{table}[ht!]
    \centering
   \begin{tabular}{cc|c|c}
Input & Neurons & RMSE & MAE \\ \hline
12    & 55       & $10.51 \pm 0.09$ & $8.28 \pm 0.13$ \\
13    & 90       & $10.58 \pm 0.11$ & $8.36 \pm 0.14$ \\
14    & 50       & $10.59 \pm 0.12$ & $8.36 \pm 0.14$ \\
14    & 45       & $10.61 \pm 0.14$ & $8.33 \pm 0.12$ \\
13    & 95       & $10.61 \pm 0.14$ & $8.39 \pm 0.11$ \\
12    & 60       & $10.61 \pm 0.14$ & $8.33 \pm 0.11$ \\
13    & 80       & $10.62 \pm 0.15$ & $8.37 \pm 0.15$ \\
13    & 85       & $10.63 \pm 0.16$ & $8.44 \pm 0.14$ \\
14    & 85       & $10.63 \pm 0.16$ & $8.35 \pm 0.15$ \\
13    & 70       & $10.63 \pm 0.16$ & $8.40 \pm 0.10$ \\
\end{tabular}
    \caption{Results for the ten best models according to the Root Mean Squared Error (RMSE) and the Mean Absolute Error on the test set for 10 independent runs.}
    \label{tab:results}
\end{table}

An intuitive initial configuration corresponds to an input 52 lags, which corresponds to a window of one year. A priori good results are expected. However, it is not pointed to as a high-quality configuration by this grid search.

\begin{figure}[!h]
    \centering
    \includegraphics[width=1.\textwidth]{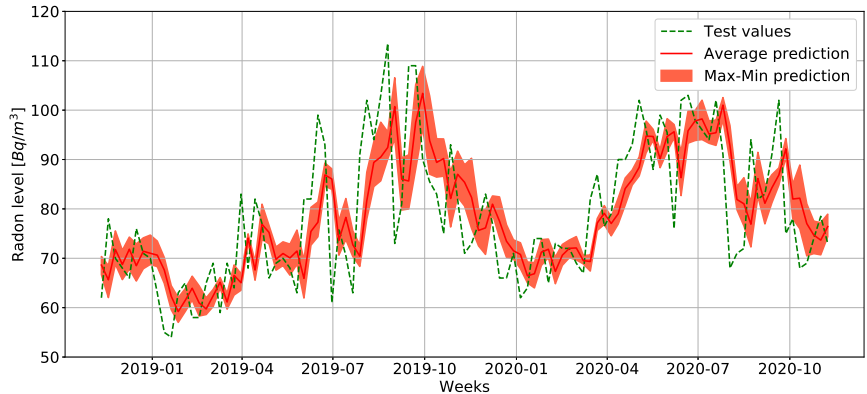}
    \caption{Comparison between the real values (green) and the average prediction (red) over the test set for the ten independent executions of the best model composed by 55 hidden neurons and input of 12 past weeks. The filled area represents the maximum and minimum values achieved for 10 independent runs.}\label{fig:pred}
\end{figure}

The radon-222 values predicted by the best model -input of 12 and 55 hidden neurons- are shown in the \ref{fig:pred} together with the real values.

It is notable that the forecasts seem to have shifted one value ahead. One might think that the network only repeated using the last value of the input window and that the rest tended to confuse the network. To compare this, the neural network performance is compared with the model $\hat{y}_t = y_{t-1}$, which essentially repeats the past value as  forecast. 

The RMSE and MAE achieved 11.42 and 8.40 respectively, and both are larger than those associated with the best neural network. Then the neural network is better than forecasting the past value. In this way, the rest of the input variables tend to improve the prediction. Indeed in as many as 67.10\% of the networks analyzed had less error.

\subsection{Variable Importance}
\begin{figure}[!h]
    \centering
    \includegraphics[width=1.\textwidth]{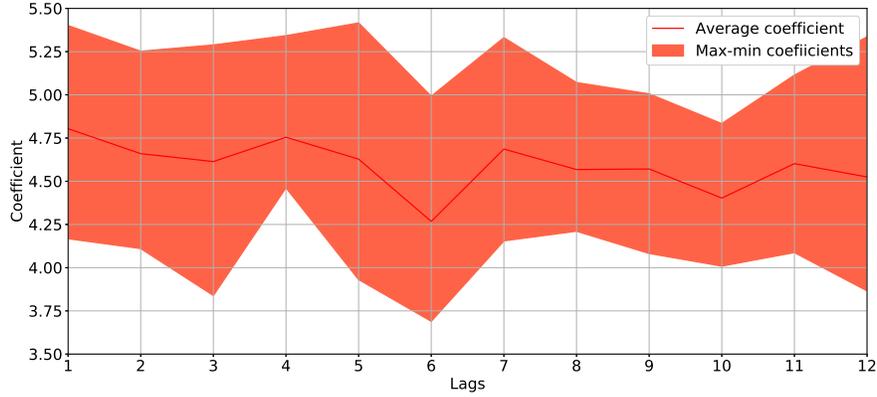}
    \caption{Garson's algorithm for the best model which is composed by 55 hidden neurons and input of 12 past weeks.The filled area represents the maximum and minimum values achieved for 10 independent runs.}\label{fig:garson}
\end{figure}

\begin{figure}[!h]
    \centering
    \includegraphics[width=1.\textwidth]{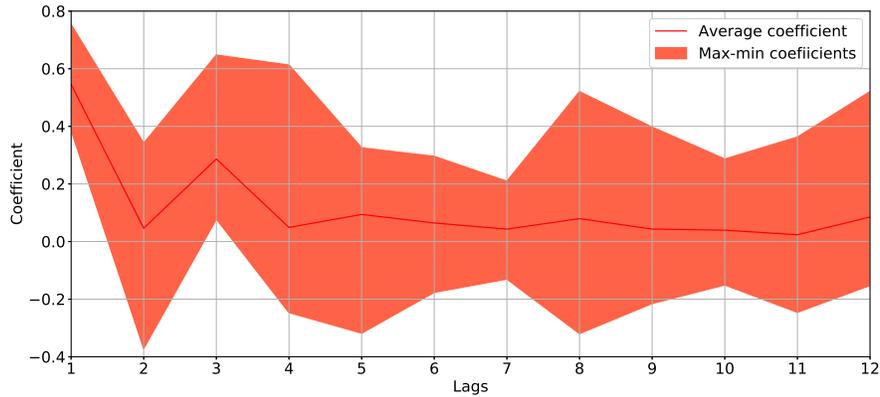}
    \caption{Olden's Algorithm for the best model which is composed by 55 hidden neurons and input of 12 past weeks.The filled area represents the maximum and minimum values achieved.}\label{fig:olden}
\end{figure}

Figures \ref{fig:garson} and \ref{fig:olden} show the coefficients of the importance of the input variables. Garson's algorithm gives a relative high importance to the first five lags, being the highest for $lag_1$. For more delayed lags, the importance reduces progressively.

In comparison with Garson's algorithm, Olden's one also gives a relatively high importance to $lag_1$, at the same time that exhibits a more critical reduction of the importance of the variables as far as the lags grow.  

\section{Conclusions}\label{section:conclusions}
In this paper a first attempt to understand the input importance in the deep models which forecast the time series of radon-222 at Canfranc Underground Laboratory is presented. A grid search was performed to find the best neural networks with a single hidden layer to forecast the radon-222 levels. The result is a neural network with input of 12 past values and 55 hidden neurons.

In order to know how the input influences the predictions, the algorithms of Olden and Garson were applied to the weights of the best model previously obtained. Both algorithms show that the $lag_1$ value has the greatest importance for the prediction. For larger lags and depending on the analysis, the relative importance goes down. Olden's algorithm still considers relevant the lags until $lag_5$. This result differs in Garson's algorithm that gives it a much lower importance. Furthermore, Olden's algorithm establishes a stronger reduction of the importance as far as the lags increase than Garson's one.


\section*{Acknowledgment}
I\~naki Rodr\'iguez-Garc\'ia is funded through the PEJ2018-003089-P project to promote Young Employment and Implementation of the Youth Guarantee in R+D+i within the framework of the State Subprogram for the Incorporation of the State Program for the Promotion of Talent and Employability in R+D+i within the framework of the State Plan for Scientific and Technical Research and Innovation 2017-2020. This contract is co-funded by the European Social Fund and the Youth Employment Initiative through  the Youth Employment Operational Program. 
Authors wish to express their thanks for the data support and sharing to Canfranc Underground Laboratory and, particularly to Iulian C. Bandac.

\end{document}